# Presentation of Low-Frequency Vibration to the Face Using Amplitude Modulation

Yuma Akiba, Shota Nakayama, Keigo Ushiyama, *Member, IEEE*, Izumi Mizoguchi, *Member, IEEE,* and Hiroyuki Kajimoto, *Member, IEEE*

***Abstract*—This study proposes a method to present pure low-frequency vibration sensations to the face that cannot be presented by small commercially available vibrators. The core innovation lies in utilizing an amplitude modulation technique with a carrier frequency of approximately 200 Hz. Due to the absence of Pacinian corpuscles in the facial region—receptors responsible for detecting high-frequency vibrations around 200 Hz—only the original low-frequency signal is perceived. Three experiments were conducted. Experiments 1 and 2 were performed on the forehead to confirm that the proposed amplitude modulation method could produce the desired low-frequency perception and to evaluate the subjective quality of the vibration. The results suggested that the proposed method could produce the perception of desired pure low-frequency vibration when applied to the forehead. In Experiment 3, the proposed method was applied to the whole face, and its range of applicability was explored. The results indicated that the original low-frequency vibration was clearly perceptible around the eyes, cheeks, and lower lip area.**

## I. INTRODUCTION

The demand for virtual reality (VR) experiences has increased with the widespread use of head-mounted displays (HMDs). However, the most widely used VR experiences lack tactile information, which is a major challenge in enhancing the sense of presence. Therefore, many studies have used wearable tactile presentation devices to enhance immersion in VR experiences [1], [2], [3]. However, wearing hand-worn devices is time-consuming in many cases. To solve this problem, attempts to integrate tactile presentation devices into HMDs have been made [4], [5], [6], [7]. This removes the additional step of putting on gloves, and current HMDs are also easier to wear.

Based on the above context, a small tactile presentation device is necessary for integration into HMDs. However, face-specific problems arise when a linear resonant actuator (LRA), which is currently the most widely used small vibrator, is employed. LRAs are small but produce strong vibrations through mechanical resonance phenomena; they can only produce vibrations around a specific resonance frequency (typically 200 Hz). On the contrary, the face is a unique part of the body, and Pacinian corpuscles, which perceive vibrations at approximately 200 Hz, do not exist [8]. Therefore, it is difficult to present clear tactile sensations to the face using small LRAs. Although some research on tactile presentation devices capable of presenting vibrations in the low-frequency range has been conducted [9], [10], these devices are not yet easily available. Currently, vibrators capable of generating low-frequency vibrations already exist. For example, the HapCoil-One (Tactile Labs) and the Haptuator [11] can generate vibrations over a wide range of frequencies. However, their relatively large size poses challenges for integration with HMDs or other face-worn interfaces.

On the other hand, a technique for presenting low-frequency sensations by modulating high-frequency vibrations has been established. Specifically, this method uses amplitude modulation to represent a low-frequency signal as an envelope wave using a carrier wave at approximately 200 Hz, which is the resonant frequency of the LRA [12], [13]. Although the envelope component can be perceived, the high-frequency component of the carrier wave is also perceptible, resulting in a noisy sensation. Therefore, the generated tactile sensation differs from a pure low-frequency sensation [14].

In this study, we propose a method for presenting a pure low-frequency sensation (i.e., only low-frequency sensation without the high-frequency component of the carrier wave) to the face using a small LRA, taking advantage of the absence of Pacinian corpuscles, a tactile feature of the face. Although the absence of Pacinian corpuscles on the face is generally a disadvantage for tactile presentation, we took advantage of this feature and used amplitude modulation to present low-frequency vibrations, as previously performed for tactile presentation on the hand. This enables the presentation of low-frequency vibrations, which is difficult for small commercially available LRAs. Even when a high-frequency wave (approximately 200 Hz) is used as the carrier wave, the face barely perceives high-frequency waves, allowing pure low-frequency vibrations to be perceived more clearly. This study aimed to confirm that amplitude modulation with a carrier wave of approximately 200 Hz, which is the typical resonant frequency of the LRA, can effectively present low-frequency vibrations of a few to several tens of hertz on the face.

Three experiments were conducted. In Experiments 1 and 2, we evaluated the frequency perception of amplitude-modulated vibrations on the forehead, forearm, and finger pad, and evaluated the quality of tactile sensation [15]. In Experiment 3, we evaluated the tactile quality of the amplitude-modulated vibrations on the entire face. All the experiments were conducted with the approval of the Ethics Committee of the University of Electro-Communications, Japan (No. H23076).

The remainder of this paper is organized as follows. Section 2 presents related work. Section 3 presents the methods of the experiments. Sections 4, 5, and 6 present the experiments conducted during the study and the respective results. Section 7 discusses the findings as well as the limitations. Finally, the conclusions are presented in Section 8.

## II. Related Work

### A. Vibration Presentation

Multiple options for haptic feedback on mobile devices are available. The most commonly used actuators are eccentric rotating mass (ERM) motors, LRAs, and piezoelectric (PZT) actuators. However, these haptic actuators have challenges, such as the inability to independently control the amplitude and frequency (ERM), and they can only vibrate near a narrow-band resonance frequency (LRA) or high frequency (PZT) [16]. Therefore, several approaches for addressing these issues have been proposed.

The first solution is to use amplitude-modulated vibration. Park et al. [13] reported sensory differences between one carrier-wave vibration and seven amplitude-modulated vibrations in psychophysical experiments. Cao et al. [17] suggested that the number of perceived intensity peaks played an important role compared to the perceived intensity waveform. Yamaguchi et al. [18] defined a human perceptual intensity model from the original signal and proposed a method of expressing the original signal by the resonance frequency of the LRA (typically 200 Hz) to transform it into an arbitrary waveform while maintaining the tactile sensation of high-frequency vibration. Waga et al. [19] used this method to maintain the tactile sensation of high-frequency vibrations to reproduce the writing comfort of pencils of different hardness. These methods focused on the alternative presentation of high-frequency vibrations and not on the presentation of low-frequency vibrations. Alma et al. [20] performed texture rendering using amplitude-modulated vibrations. While these methods can perceive the frequency of the envelope wave, the sensation is noisy because the high-frequency component of the carrier wave is also perceived, and the tactile sensation is different from the original pure low-frequency sensation [14].

The second solution involves methods that exploit the hardware features. Liu et al. [9] achieved a strong Lorentz force by placing a magnetic disk between the copper coils. Manabe et al. [10] developed a structure in which repelling permanent magnets are bonded together such that the generated leakage flux is orthogonal to the coil, enabling efficient low-frequency vibration presentation in a compact size; however, they are still in the development stage and not easily available. Additionally, Dosen et al. [21] proposed a device that can simultaneously change the amplitude and frequency of vibrations by inputting a short-pulse waveform into a large brushless DC motor to generate complex vibrotactile stimuli. Yem et al. [22] proposed a method for producing low-frequency vibrations using a DC motor. However, both motors need to be large to present low-frequency vibration and are not suitable for integration into HMDs.

### B. Haptics on the Face

The tactile facial presentation has been extensively employed in the research and development of welfare devices, which are designed to convert visual information into tactile feedback for individuals with visual impairments; for instance, Kajimoto et al. [23] introduced a device that conveys visual cues to the face through tactile presentation with electrical stimuli, thereby enabling visually impaired users to perceive essential visual information.

Recently, with the widespread use of HMDs, many efforts to incorporate tactile presentation devices into HMDs to add tactile information to the VR experience have been made. Chang et al. [24] proposed the use of servomotors to tighten the belt portion of an HMD to present pure pressure on the face. Shen et al. [6] mounted ultrasound presentation arrays at the bottom of an HMD and converged ultrasound waves to present tactile sensations to the face. Gugenheimer et al. [25] incorporated a gyro-effect force-presentation mechanism using a flywheel in an HMD to present force sensations. However, the size of these devices limits their wearability.

Conversely, many proposals have been made for devices that incorporate a tactile sensation presentation device into HMDs in a compact size. Tseng et al. [26] attached an air jet nozzle inside an HMD to present a sense of contact with the air pressure around the eyes. Peiris et al. [5] created an HMD with built-in Peltier elements for temperature sensory presentation and produced content using directional perception and temperature presentation. Oliveia et al. [27] incorporated vibrators into the cushion of an HMD and verified their effectiveness for directional perception in a VR space. Similarly, Valkov et al. [28] built a vibrator into the cushion of an HMD and presented vibrations to the face. Their system enhances VR safety by providing information about the distance to real-world objects, rather than VR elements. Chu et al. [29] developed an HMD with multiple vibrators attached to the side of the head and demonstrated the effectiveness of the vibrators in teaching directions in a VR environment and in presenting information between user movements. However, none of them presented pure low-frequency sensations in a compact format. “A compact format” in this context refers to a size suitable for integration into HMDs or mask-type interfaces. Specifically, this corresponds to a form factor of approximately 10 mm or less, making implementation feasible within such wearable systems.

## III. Methods

This section describes the equipment and methods common to all experiments. Specifically, it provides details on the experimental apparatus, including the actuator, driver, and amplitude modulation method.

The experimental devices are shown in Fig. 1a. The LRA was attached to the center of the forehead, to the ventral part of the forearm, and to the index finger pad for Experiments 1 and 2, as illustrated in Fig. 1b. The choice of the index finger pad was based on its common usage for tactile presentation because of its high spatial resolution for tactile sensation. The forearm was selected as a site where Pacinian corpuscles are located, but with a relatively lower spatial resolution than the index finger pad. The forearm was chosen to investigate whether the results of the forehead experiment stemmed from low tactile receptor

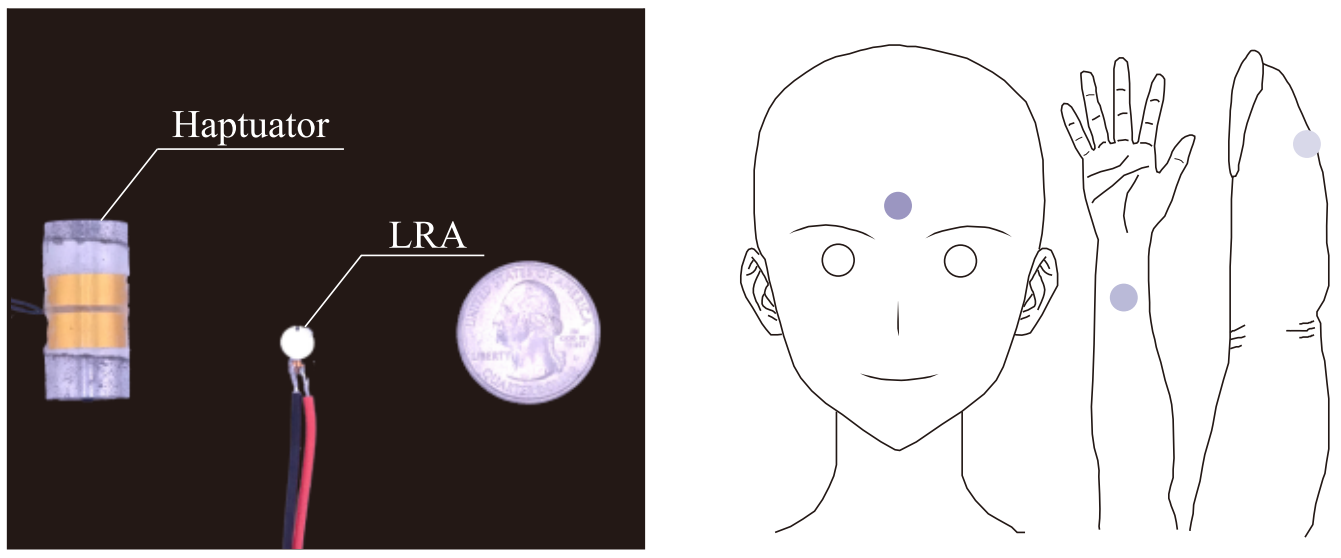

(a) (b)

Fig. 1. Experimental conditions: (a) Apparatus used in experiments: TL002 14-A (Tactile Labs), and VG0640001D (Vybronics). (b) Body site, center of the forehead, ventral part of the forearm, and index finger pad.

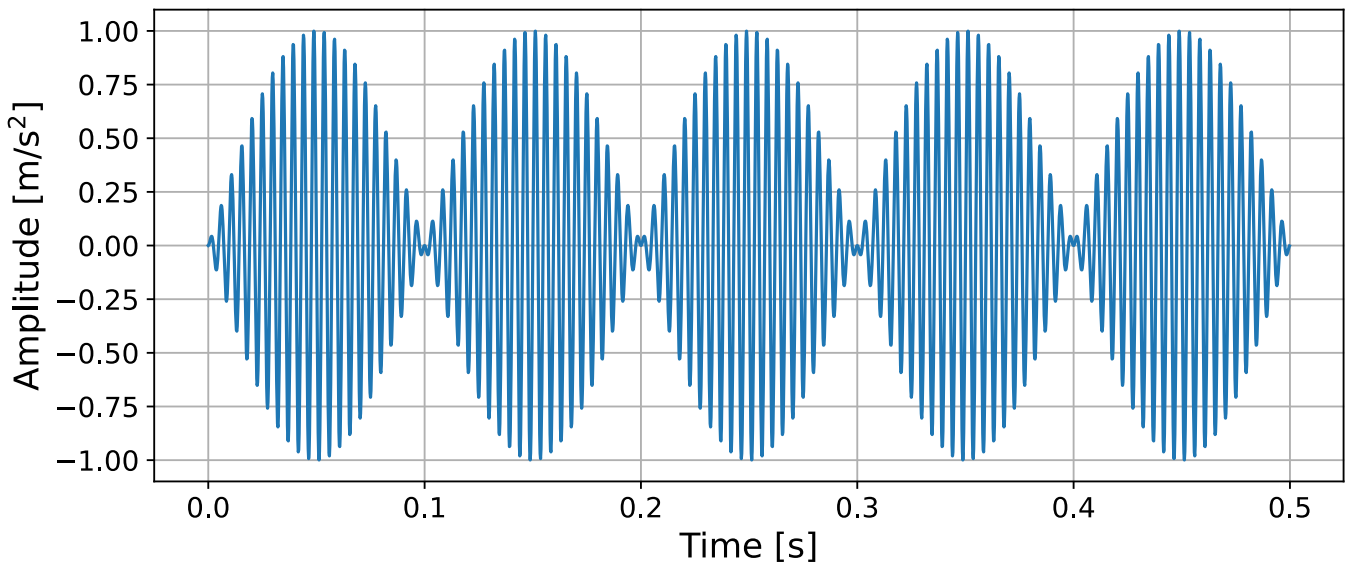

Fig. 2. A representation of the amplitude-modulated stimulus in equation (1). This stimulus represents a wave with a carrier wave of 210 Hz and an envelope wave of 10 Hz. Amplitude-modulated vibrations were presented using LRA.

density or the absence of Pacinian corpuscles.

The device producing the amplitude-modulated vibration signals was a small LRA (VG0640001D, Vybronics). The size of this device was 6 mm in diameter. The amplitude-modulated vibration signals were generated using a function generator (DG812, RIGOL) and amplified using an audio amplifier. The amplitude modulation wave used in this study is expressed according to the following equation:

$$g\ t\ = A \sin\ 2\pi f_e t\ \sin\ 2\pi f_c t \qquad 1$$

where $f_e$ represents the frequency of the original signal. $f_c$ represents the carrier frequency of the amplitude modulation, and the resonant frequency of the LRA (210 Hz) was used. Through amplitude modulation, the frequency of the original signal is theoretically altered by the carrier wave, and signal $g(t)$ contains frequencies $f_c \pm f_e$. The representation of the amplitude-modulated stimulus is presented in Fig. 2. Accordingly, the modulation signal with $f_e$ results in an envelope wave that contains a frequency component at twice the modulation frequency—that is, $2f_e$. For example, a 4 Hz modulation signal generates an envelope with a repetition frequency of 8 Hz, and therefore a perceivable 8 Hz beat. In this paper, the repetition frequency of the envelope wave ($2f_e$) will hereafter be referred to as the frequency of the envelope wave.

Before conducting the experiments, we confirmed whether the vibration actuators, LRA and Haptuator [11] (TL002-14-A, Tactile Labs, Fig. 1a.), could generate a perceivable vibration. The Haptuator was used only in Experiment 1 to compare the sinusoidal and amplitude-modulated vibration. One author measured the vibrations using an accelerometer (BMX055, BOSCH). For the measurement, the LRA was driven by an amplitude-modulated wave signal, whereas the Haptuator was driven by a sinusoidal wave signal. Sinusoidal wave signals were generated using a custom program (Processing), and the signals were input to the vibration actuator through an audio interface (UltraLite mk5, MOTU) and a stereo audio amplifier (FX-AUDIO-, FX202A/FX-36A PRO, North Flat Japan). The vibration amplitudes were measured with a constant signal amplitude, and the participants could clearly perceive the vibrations at 2, 4, 8, 16, and 32 Hz. Measurements were conducted at the same body locations as in Experiments 1 and 2 (see Table I). The actuator was attached with surgical tape (Transpore, 3 M) in the same way as in all experiments. The acceleration data was obtained by taking the maximum value of the acceleration of the vibration for 5 seconds 5 times, and the average value was used as the acceleration data.

The results are illustrated in Fig. 3. They indicated almost consistent acceleration at the forehead, forearm, and index finger pad. The Haptuator at 2 Hz has its own actuator limit, and the acceleration decreases. Based on this result, the amplifier volume used for the experiments remained constant across all participants throughout Experiments 1 and 2 and was not individually adjusted. The acceleration of the LRA in Experiment 3 was set to 40 m/s$^2$.

## IV. Experiment 1

In this experiment, participants were asked to match the frequency of the envelope wave from the amplitude-modulated vibration—presented on the forehead, forearm, and finger pad of the dominant hand—with the frequency of the sinusoidal vibrations applied to the finger pad of the non-dominant hand. This confirmed that the low-frequency component of the envelope wave on the forehead could be perceived. We also tested whether the absence of the Pacinian corpuscles in the forehead allowed for more accurate frequency matching compared to other areas with the Pacinian corpuscles.

### *A. Procedure*

We recruited 11 participants for this experiment (ten males and one female, aged 22–24 years from our university). To compare the sinusoidal and amplitude-modulated vibration signals, we used the Haptuator to represent simple sinusoidal vibrations. Sinusoidal wave signals were generated using the same method as in the acceleration measurement setup.

We asked the participants to adjust the frequency of the sinusoidal vibration using the Haptutor on the non-dominant hand to match the frequency of vibration perceived in the other location. We presented amplitude-modulated vibrations with a small LRA at the index finger pad on the dominant hand side (Finger condition), forearm on the dominant hand side (Arm condition), and forehead (Head condition). The LRA was secured to the body using surgical tape. We also prepared a condition in which sinusoidal vibrations were presented using the Haptuator, which was grasped with the index finger and thumb on the dominant hand side (Haptuator condition). The Haptuator condition served as a control group to establish a baseline for matching perception. Therefore, two Haptuators (i.e., one in the dominant hand and one in the non-dominant hand) were used in the Haptuator condition. A summary of the conditions is presented in Table I and Fig. 4.

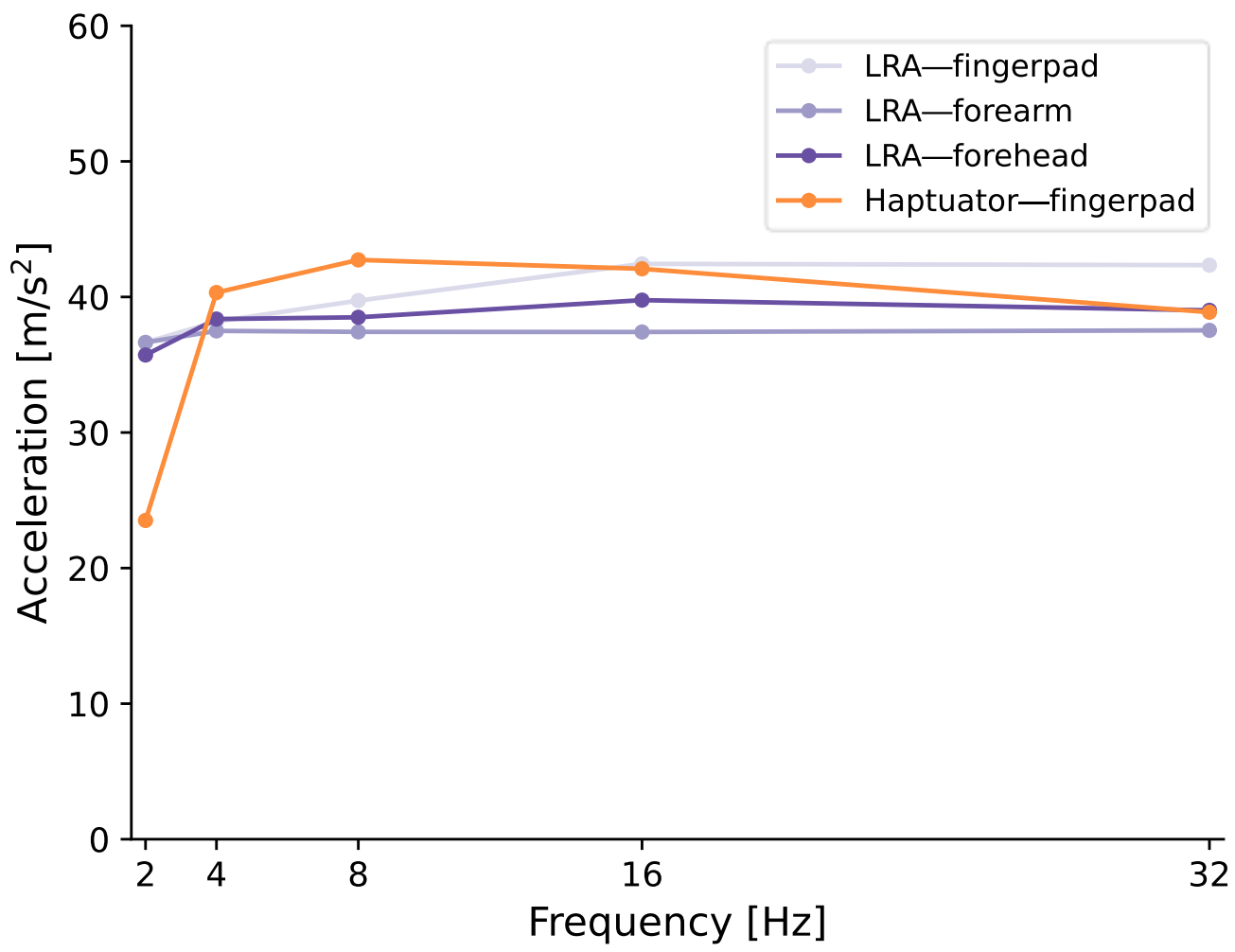

Fig. 3. Result of acceleration measurements at each frequency used in two experiments. Amplitude modulation signals were applied to LRA, whereas sinusoidal signals were applied to the Haptuator.

TABLE I
EXPERIMENTAL CONDITIONS

| Condition | Vibrator | Body site | Signal | Frequency [Hz] |
|---|---|---|---|---|
| Finger | LRA | Finger pad | AM wave | 2,4,8,16,32 |
| Arm | LRA | Forearm | AM wave | 2,4,8,16,32 |
| Head | LRA | Forehead | AM wave | 2,4,8,16,32 |
| Haptuator | Haptuator | Figer pad | Sine wave | 2,4,8,16,32 |

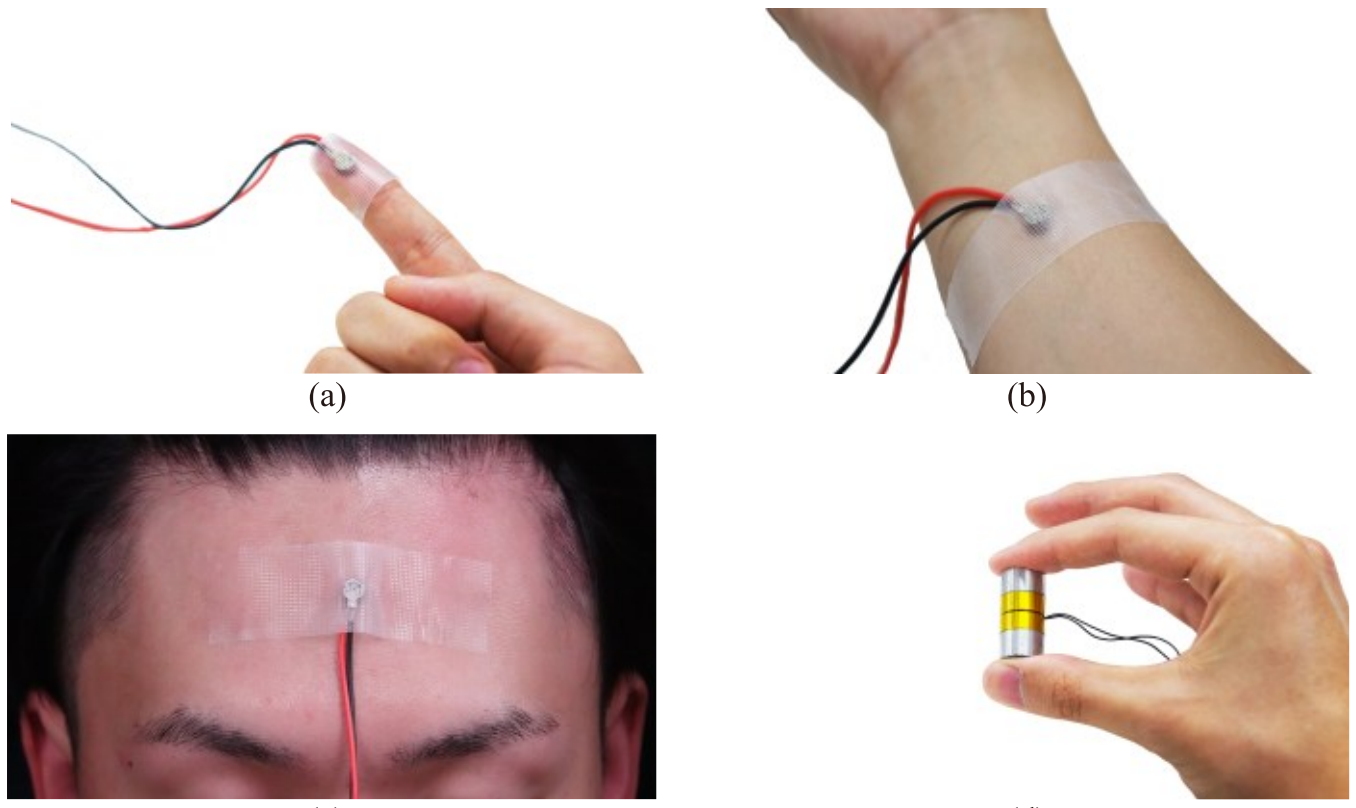

(a) (b) (c) (d)

Fig. 4. Summary figure corresponding to Table I. (a) Finger condition, (b) Arm condition, (c) Head condition, and (d) Haptuator condition. The LRA was secured to the body with surgical tapes (Transpore, 3M). The Haptuator was grasped with the index finger and thumb. This shows an example of a right-handed participant. In all cases, participants grasp a Haptuator in their non-dominant hand, and its frequency was adjusted.

The frequency was adjusted by pressing the arrow keys on the keyboard using fingers other than the thumb and index finger of the non-dominant hand, as illustrated in Fig. 5. Pressing the up-arrow key increased the frequency by 1 Hz, whereas pressing the down-arrow key decreased the frequency by 1 Hz. When the desired frequency was reached, the user finalized it by pressing the Enter key. The participants regulated their visual focus by keeping their eyes closed, except when it was necessary to locate or manipulate keys. They wore headphones and listened to white noise to mask auditory cues.

Five stimulation frequencies were used: 2, 4, 8, 16, and 32 Hz (as noted in the previous section, 1, 2, 4, 8, and 16 Hz sinusoidal waves were used to generate 2, 4, 8, 16, and 32 Hz envelope waves. Unless otherwise specified, the stimulus frequency refers to the repetition frequency of the envelope wave). They were modulated by a 210-Hz (resonant frequency of the LRA) carrier wave to drive the LRA. The duration of stimulation was not limited, and stimulation continued until the participants finished adjusting their frequency. The adjustment started at 1 Hz or 300 Hz for each trial. The order of the stimulation frequency (2, 4, 8, 16, and 32 Hz) and the starting frequency (1 or 300 Hz) were randomized. The order of the conditions (Finger, Arm, Head, and Haptuator) was also randomized among participants. The experiment employed a within-subjects design, with all participants completing all four conditions (Finger, Arm, Head, and Haptuator). Each condition involved five types of frequencies (2, 4, 8, 16, and 32 Hz) and two types of starting frequencies (1 Hz and 300 Hz), with two trials conducted for each combination. Therefore, the total number of trials per participant was 80 (4 conditions × 5 frequencies × 2 starting frequencies × 2 trials). There were no limits to the adjustment time.

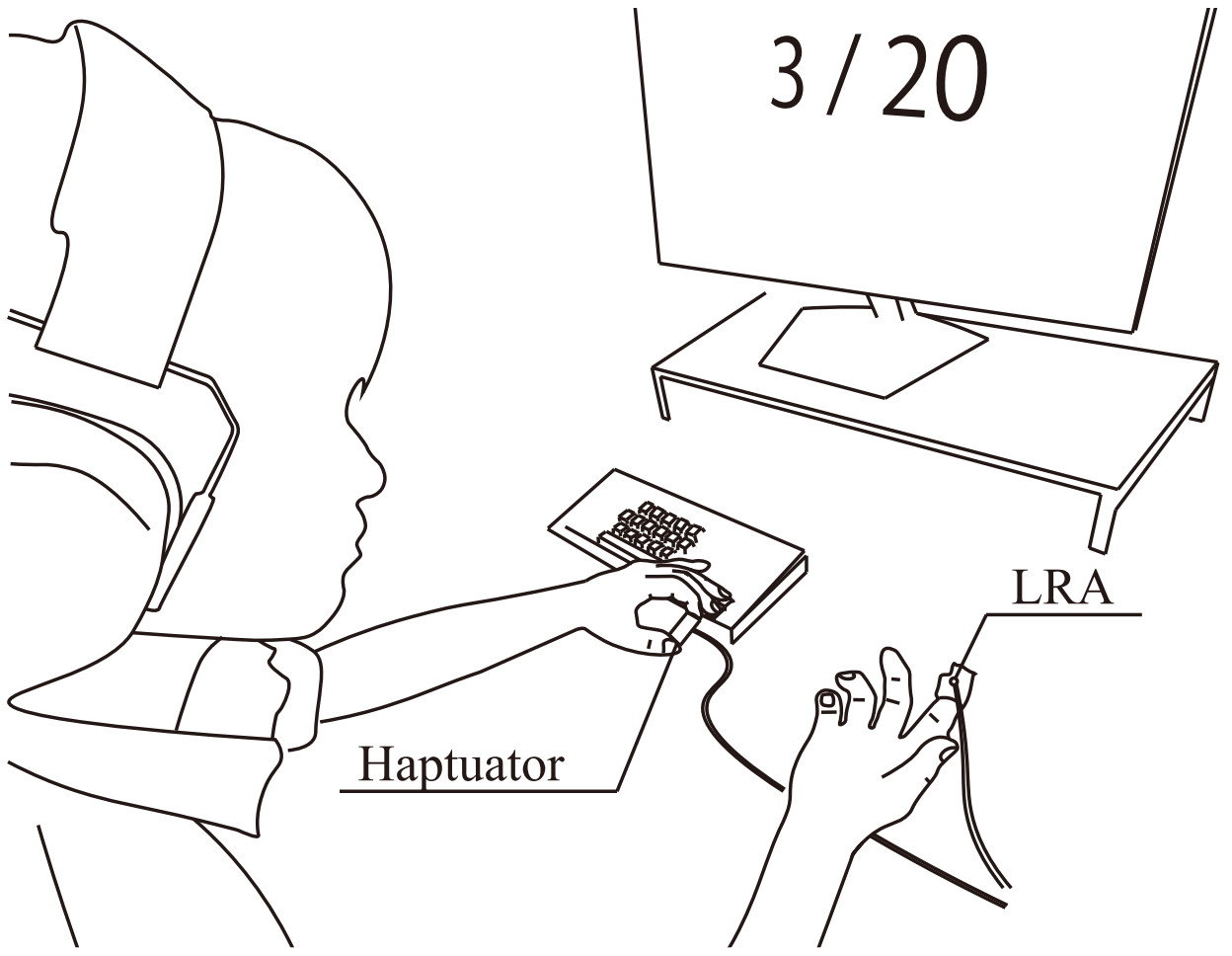

Fig. 5. Example scene of the experiment with a right-handed participant (Finger condition). The monitor showed the number of remaining trials; however, the participants were instructed to keep their eyes closed except when operating the keys as much as possible.

Prior to the measurement at each stimulation frequency, the participants conducted two practice trials in the same manner as the measurement. At the end of the experiment, the participants provided open comments about the experiment in general. The participants were not given feedback on the accuracy of their adjustments.

*B. Results*

The perceived frequency results for each set frequency and site are presented in Fig. 6. The frequency value for each participant was determined as the average of the values obtained from two trials at a starting frequency of 300 Hz and two trials at a starting frequency of 1 Hz, covering the range from 1 Hz to 300 Hz. Additionally, Fig. 7 presents the data from trials in which the stimulation started at 1 Hz, while Fig. 8 shows the data from trials in which the stimulation started at 300 Hz. The perceived frequency for each participant was calculated as the average of two trials with the same starting frequency.

We first applied an Aligned Rank Transform (ART) ANOVA to the data. This non-parametric method allows for testing the main effects and interactions of two factors with non-normally distributed data. Specifically, we examined the effects of body site (Finger, Arm, Head, Haptuator) and starting frequency (1 Hz vs. 300 Hz) as two independent factors.

The results revealed that, for the 2 Hz condition, neither the body site [$F(3, 80) = 0.7$, $p = 0.545$] nor the starting frequency [$F(1, 80) = 0.1$, $p = 0.740$] showed significant main effects, and no interaction effect was found between the two factors [$F(3, 80) = 0.1$, $p = 0.979$]. For the 4 Hz condition, significant main effects were found for both body site [$F(3, 80) = 8.5$, $p < 0.001$] and starting frequency [$F(1, 80) = 12.4$, $p < 0.001$], with no significant interaction effect [$F(3, 80) = 1.1$, $p = 0.344$]. For the 8 Hz condition, significant main effects were found for both body site [$F(3, 80) = 3.5$, $p = 0.020$] and starting frequency [$F(1, 80) = 9.0$, $p = 0.004$], with no significant interaction effect [$F(3, 80) = 0.4$, $p = 0.750$]. For the 16 Hz condition, significant main effects were found for both body site [$F(3, 80) = 7.6$, $p < 0.001$] and starting frequency [$F(1, 80) = 22.6$, $p < 0.001$], with no significant interaction effect [$F(3, 80) = 1.9$, $p = 0.138$]. For the 32 Hz condition, significant main effects were found for both body site [$F(3, 80) = 5.4$, $p = 0.002$] and starting frequency [$F(1, 80) = 21.5$, $p < 0.001$], with no significant interaction effect [$F(3, 80) = 1.5$, $p = 0.209$].

For each stimulation frequency condition (2, 4, 8, 16, and 32 Hz), the Friedman test was conducted for all data. The test indicated the main effects of the 4 Hz ($p < 0.001$), 8 Hz ($p < 0.01$), and 16 Hz ($p < 0.001$) conditions. Multiple comparisons with the Bonferroni correction revealed 5% significant differences for the Arm and Head conditions at 4 Hz ($p = 0.046$), Arm and Head conditions at 16 Hz ($p = 0.023$), and Arm and Haptuator conditions at 16 Hz ($p = 0.035$).

A similar analysis was separately conducted for data from trials starting at 1 Hz and 300 Hz. For data from trials starting at 1 Hz, the Friedman test indicated the main effects of the 8 Hz ($p < 0.05$) and 16 Hz ($p < 0.05$) conditions. Multiple comparisons with the Bonferroni correction revealed 5% significant differences for the Arm and Head conditions at 8 Hz ($p = 0.044$), Arm and Haptuator conditions at 8 Hz ($p = 0.035$), Arm and Head conditions at 16 Hz ($p = 0.04$), and Arm and Haptuator conditions at 16 Hz ($p = 0.048$). For data from trials starting at 300 Hz, the Friedman test indicated the main effects of the 4 Hz ($p < 0.01$), 8 Hz ($p < 0.01$), and 16 Hz ($p < 0.001$) conditions. Multiple comparisons with the Bonferroni correction revealed 5% significant differences for the Arm and Head conditions at 8 Hz ($p = 0.035$), Arm and Finger conditions at 16 Hz ($p = 0.029$), Arm and Head conditions at 16 Hz ($p = 0.006$), and Arm and Haptuator conditions at 16 Hz ($p = 0.041$). Multiple comparisons with the Bonferroni correction did not reveal significant differences for the 4 Hz condition.

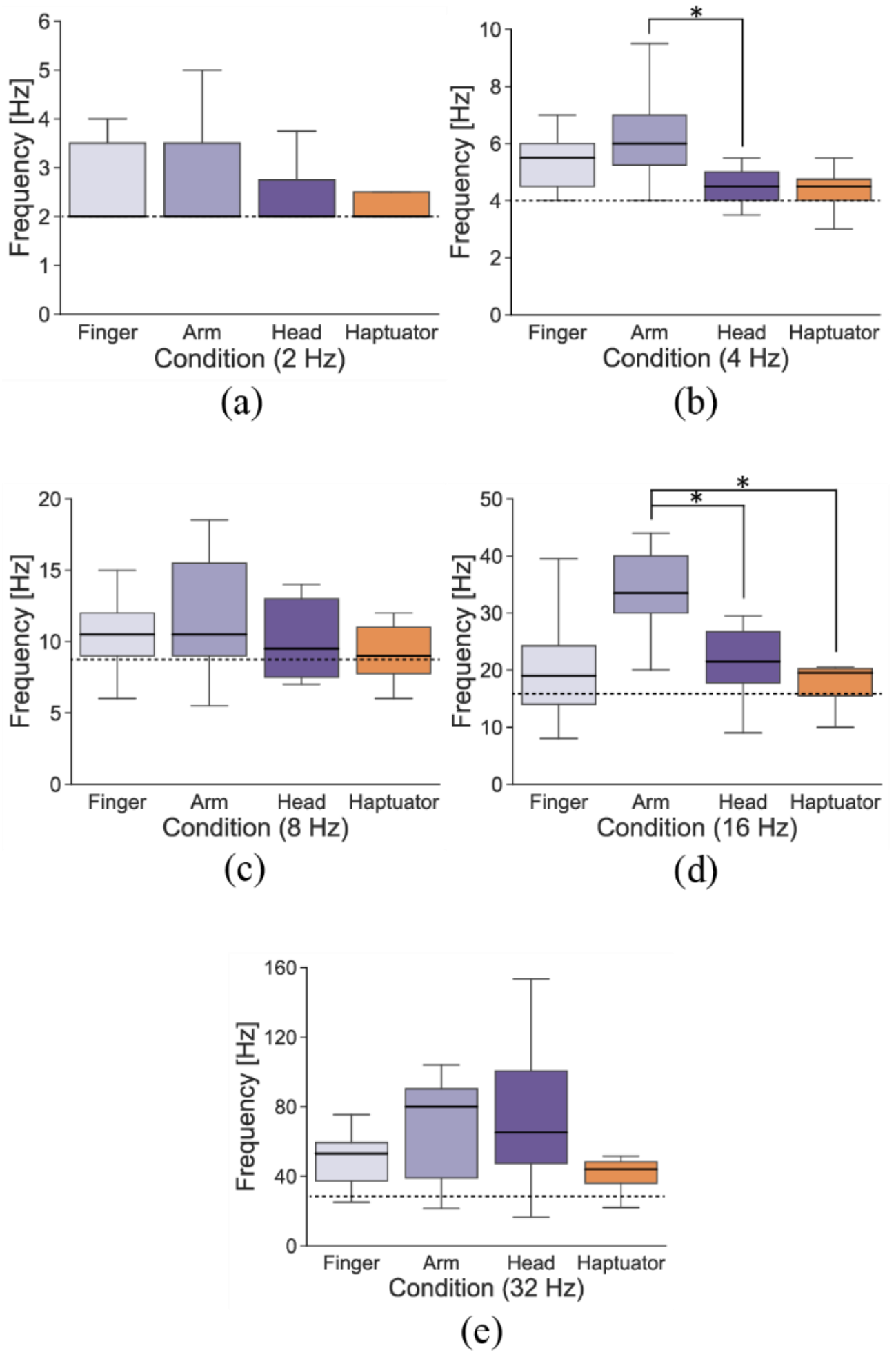

Fig. 6. Averaged perceived frequency across all data. The dotted lines indicate the reference vibrotactile stimuli. The purple-colored graphs represent the group with amplitude-modulated vibration using LRA, while the orange one represents the baseline condition with sinusoidal vibration. (a) 2 Hz, (b) 4 Hz, (c) 8 Hz, (d) 16 Hz, and (e) 32 Hz (*: $p < 0.05$)

Following the experiment, participants made comments such as "Lower frequencies were easier to match," "Matching frequencies with the forehead was easier than with the forearm," and "Matching frequencies with the Haptuator (sinusoidal wave) was the easiest". Some also commented, "The task of adjusting frequencies itself was challenging," and "Starting with a higher trial frequency made it harder to match than starting with a lower frequency."

## V. Experiment 2

In this experiment, we examined how people perceived the quality of sensations of amplitude-modulated vibrations on their index finger pad, forearm, and forehead. The purpose of this experiment was to assess whether the participants perceived pure low-frequency vibrations.

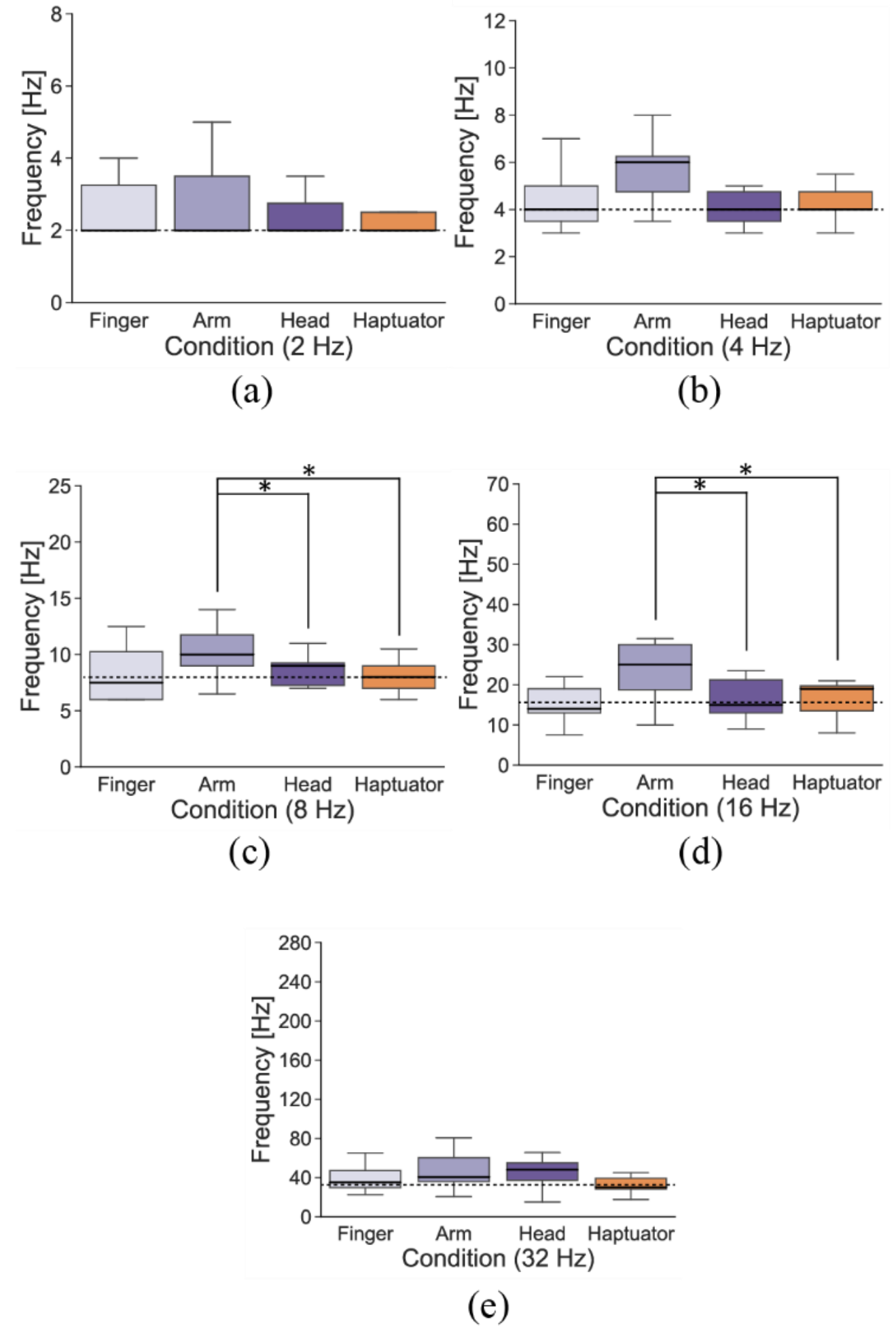

Fig. 7. Average perceived frequency for trials with starting frequency of 1 Hz. The dotted lines indicate the reference vibrotactile stimuli. The purple-colored graphs represent the group with amplitude-modulated vibration using LRA, while the orange one represents the baseline condition with sinusoidal vibration. (a) 2 Hz, (b) 4 Hz, (c) 8 Hz, (d) 16 Hz, and (e) 32 Hz (*: $p < 0.05$)

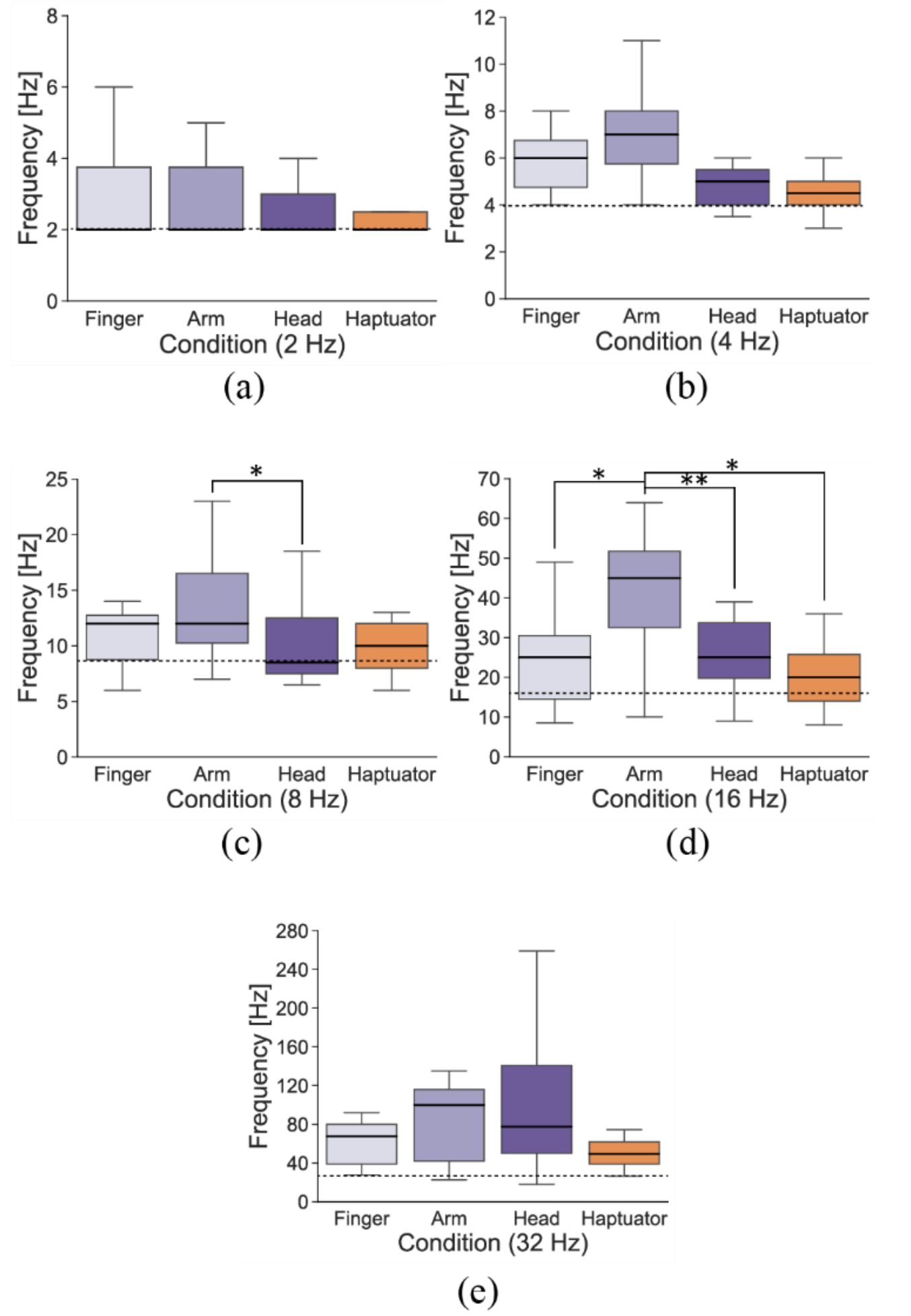

Fig. 8. Average perceived frequency for trials with starting frequency of 300 Hz. The dotted lines indicate the reference vibrotactile stimuli. The purple-colored graphs represent the group with amplitude-modulated vibration using LRA, while the orange one represents the baseline condition with sinusoidal vibration. (a) 2 Hz, (b) 4 Hz, (c) 8 Hz, (d) 16 Hz, and (e) 32 Hz (*: $p < 0.05$, **: $p < 0.01$)

### *A. Procedure*

We recruited ten participants (ten males, aged 22–24 years) in this experiment (eight of them also participated in Experiment 1). We applied the same vibratory stimulation as in Experiment 1 (i.e., Finger, Arm, and Head conditions in Table I). Five amplitude-modulated vibration conditions were used: 2, 4, 8, 16, and 32 Hz. The carrier frequency was 210 Hz. The duration of stimulation was 10 s. The experiment was conducted using a within-subjects design, with all participants completing all three conditions (Finger, Arm, and Head). In each condition, five types of frequencies (2, 4, 8, 16, and 32 Hz) were presented, and one trial was conducted for each frequency. Therefore, the total number of trials per participant was 15 (3 conditions × 5 frequencies).

The participants then evaluated the quality of tactile sensations using a 9-point Likert scale from 1 to 9 for the following attributes: (A) speed (slow-fast), (B) smoothness (bumpy-smooth), (C) hardness (soft-hard), (D) strength (weak-strong), (E) distinctness (vague-distinct), (F) weight (light-heavy), (G) clarity (dull-clear), and (H) preference (dislike-like) of vibration, based on previous studies [30], [31]. The participants wore headphones and listened to white noise to mask the auditory cues. The order of the amplitude modulation frequency and presenting body site was randomized. At the end of the experiment, the participants provided open comments about their sensations and the experiment.

### *B. Results*

The results are presented in Fig. 9. It depicts the answers regarding the quality of vibrations at different frequencies and sites. The Friedman test revealed the main effects of distinctness ($p < 0.05$) and clarity ($p < 0.01$) at 2 Hz; distinctness ($p < 0.05$) and clarity ($p < 0.05$) at 4 Hz; distinctness ($p < 0.001$) and clarity ($p < 0.001$) at 8 Hz; and smoothness ($p < 0.01$), distinctness ($p < 0.01$), and clarity ($p < 0.001$) at 16 Hz. For multiple comparisons with Bonferroni correction, 5% significant differences were observed for Arm and Head at 2-Hz distinctness ($p = 0.041$), Arm and Head at 2-Hz clarity ($p = 0.026$), Arm and Head at 4-Hz distinctness ($p = 0.044$), Finger and Head ($p = 0.017$), Arm and Head at 4-Hz

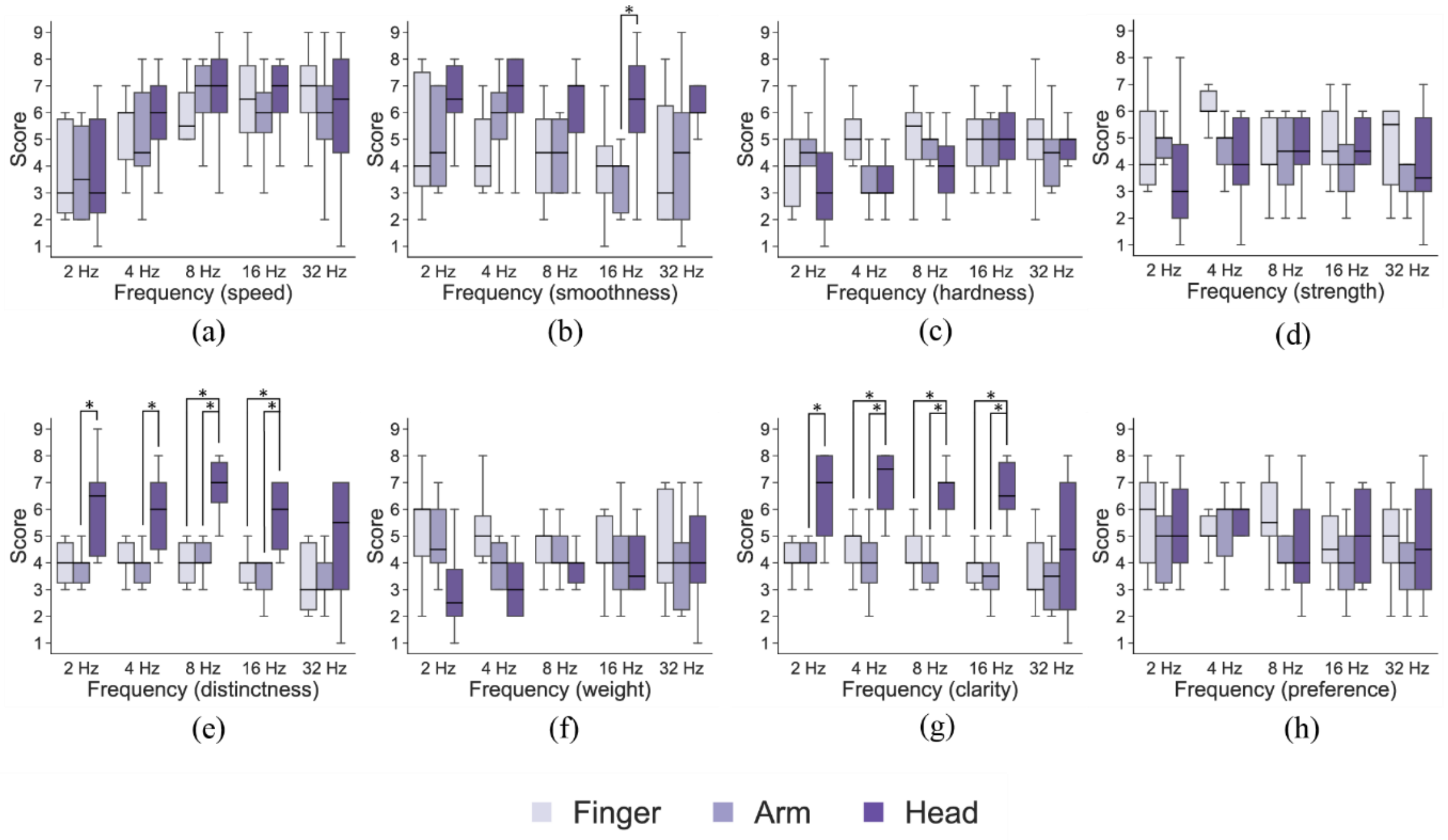

Fig. 9. Results of vibration quality responding with adjective pairs. (a) A: speed (slow-fast), (b) B: smoothness (bumpy-smooth), (c) C: hardness (soft hard), (d) D: strength (weak-strong), (e) E: distinctness (vague-distinct), (f) F: weight (light-heavy), (g) G: clarity (dull-clear), and (h) H: preference (dislike-like) (*: $p < 0.05$)

clarity ($p = 0.026$), Finger and Head ($p = 0.026$), Arm and Head at 8-Hz distinctness ($p = 0.017$), Finger and Head ($p = 0.036$), Arm and Head at 8-Hz clarity ($p = 0.025$), Arm and Head at 16-Hz smoothness ($p = 0.048$), Finger and Head ($p = 0.03$), Arm and Head at 16-Hz distinctness ($p = 0.039$), and Finger and Head ($p = 0.026$), Arm and Head at 16-Hz clarity ($p = 0.026$).

Following the experiment, participants made comments such as "The tactile sensation felt on the forehead seemed quite different from that on the forearm and finger pad." Additionally, they remarked, "On the forehead, it felt softer and smoother," and they perceived, "The higher frequency felt faster in speed." They found that "The finger pad and forearm had similar vibration sensations."

## VI. Experiment 3

In this experiment, we identified areas on the face where low-frequency sensations could be effectively presented, with future applications in mind, such as integrating vibrators into HMDs, masks, and smart glasses.

### *A. Procedure*

We used the entire face to identify areas that perceived pure low-frequency sensations through amplitude-modulated vibrations, including the forehead. We recruited 10 participants (eight males and two females, aged 21–24 years) in this experiment. We applied LRA to 16 points on the face, index finger of the dominant hand, and forearm of the dominant hand using surgical tape (Transpore, 3M). LRA was also applied to two points above the ear, one point on the top of the head, and one point on the back of the head using Velcro tape wrapped around the head.

The detailed stimulus points are shown in Fig. 10. Point pb was set 1 cm above the eyebrows and just above the nose. Points p1 and p2 were set so that "the distance from ear to ear": "the distance from ear to p1 and p2" = 5:2. Points p3 and p4 were set so that "the distance from the ear to the eye": "the distance from the eye to p3 and p4" = 9:1. Points p5 and p6 were set directly under the eyes with "the distance from the eyes to the chin": "the distance from the eyes to p5 and p6" = 11:1. Points p7 and p8 were set directly under the eyes with "the distance from the eyes to the chin": "the distance from the eyes to p7 and p8" = 11:4. Points p9 and p10 were set so that "distance from under the ear to the mouth": "distance from mouth to p9 and p10" = 13:2. Points p11 and p12 were set at the midpoint between the edge and middle of the lips. Point p13 was located in the middle of the nose and mouth. Point p14 was set at the center of the nose. Point p15 was set at the middle of the mouth and chin. Point p16 was set at the tip of the chin. Point p17 was set at the top of the head. Points p18 and p19 were set 1 cm away from the top of the ear. Point p20 was set at the back of the head. Point p21 was set at the ventral part of the forearm. Point p22 was set at the index finger pad. Each point was determined using a measuring tape and marked in advance using a water-based pen. The pressing force of the vibrator was measured using a Shocak Chip 6DoF-P18 (Touchence) and adjusted to approximately 0.5 N.

The carrier frequency was set to 210 Hz, which is the resonant frequency of the LRA, and the original signal was amplitude-modulated (10-Hz envelope wave). The acceleration of the LRA in Experiment 3 was set to 40 $m/s^2$. The participants were asked to report the quality of the tactile sensations they experienced during vibration. In the evaluation method, amplitude-modulated vibration was first presented to the forehead (Point pb) for 10 s. After a 1-s interval, a 10-s amplitude-modulated vibration was presented at one of the 22 points. Thus, the participant received the vibration on point pb before each stimulation at the other points. The participants rated the similarity of the tactile sensation to that of the forehead on a 9-point Likert scale (1 = not similar, 9 = similar). The experiment was conducted using a within-subjects design, with all participants completing all 22 stimulation conditions. Each stimulation site was tested three times in a random order. Therefore, the total number of trials per participant was 66 (22 conditions × 3 trials). The participants wore headphones during the experiment, and white noise masked the auditory cues.

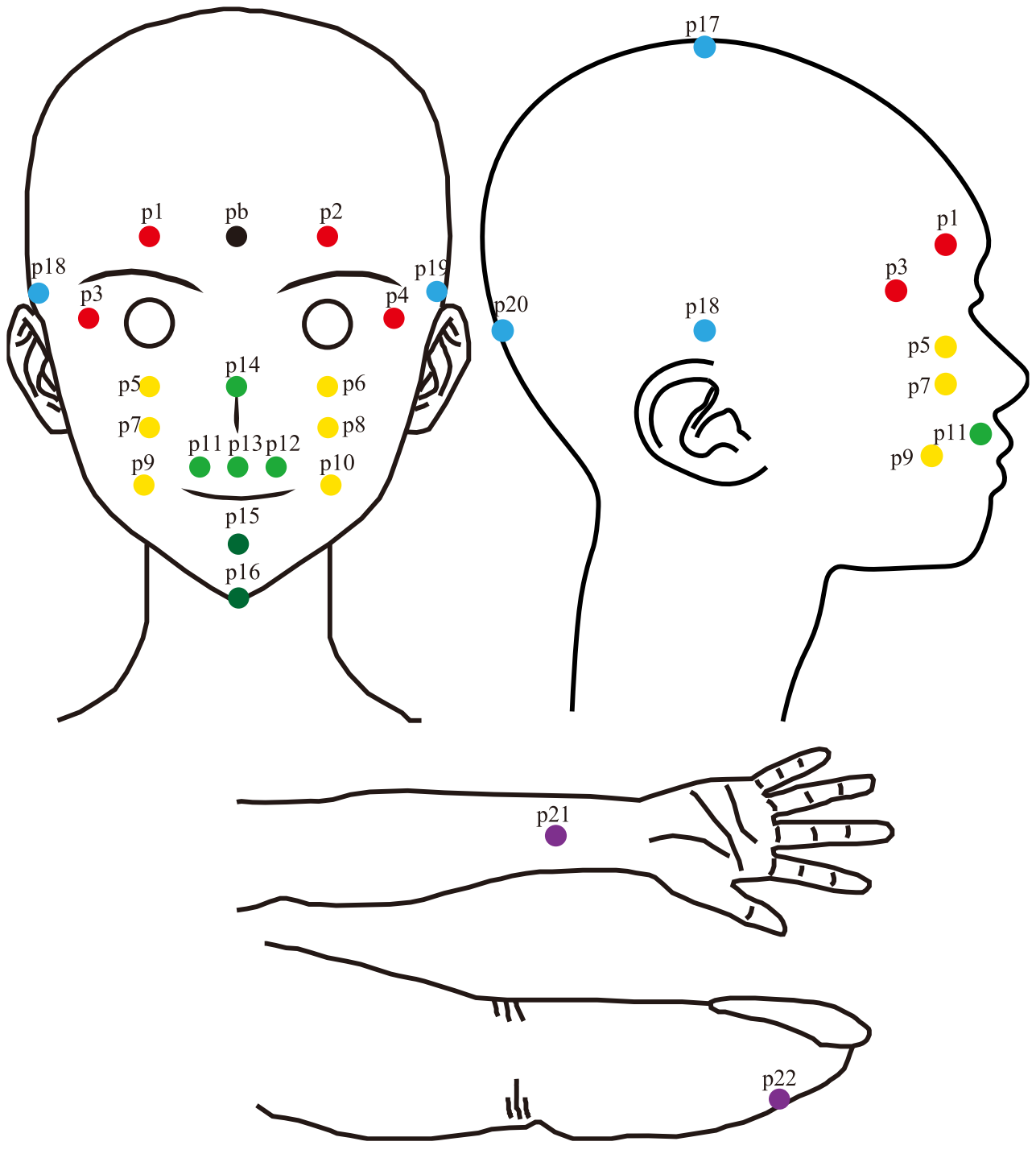

Fig. 10. Stimulus point diagram. 22 stimulation points were selected: 16 points on the face, 2 points on the upper ears, 1 point on the top of the head, 1 point on the back of the head, 1 point on the index finger of the dominant hand, and 1 point on the forearm of the dominant hand. The red group represents the temples and forehead. The yellow group represents the area from under the eyes to the cheeks. The light green-yellow group represents the area from the nose to the upper part of the mouth. The green group represents the lower part of the mouth and chin. The light blue group represents the areas around the head (top of the head, sides of the head, and back of the head). The purple group represents the finger pad and forearm.

### *B. Results*

The similarity assessment scores are shown in Fig. 11. Point pb was set at 9, and comparisons were made with other areas (p1 to p22) using the Wilcoxon signed-rank test with Bonferroni correction. Since there were a large number of areas to compare, no significant differences were detected. However, an examination of the graph reveals a clear trend indicating a division into two groups based on whether the median is below or above 5. As for the padded area where the HMD was worn (around the eyes), cheeks, and points were the areas where the median value was above 5 and elicited a sensation similar to that of point pb. On the other hand, the nose, upper lip, temporal region, and parietal region were the areas where the median was below 5 and elicited a sensation distinct from that of point pb (i.e., a perception with an accompanying noise component).

## VII. Discussion

### *A. Frequency Perception*

Experiment 1 showed that the accuracy of perceiving vibrations on the index finger pad and forehead was similar. It aligned with the result of the previous study using the finger pad, where participants could discriminate the envelope of the amplitude-modulated wave when the signal wave was below 50 Hz [32].

Overall, the adjusted frequencies were slightly higher than the envelope wave repetition frequencies. For example, a 4-Hz amplitude-modulated envelope wave on the finger pad and forearm was matched by a 5–6 Hz sinusoidal wave, and an 8-Hz amplitude-modulated envelope wave on the finger pad and forearm was matched by a 10-Hz sinusoidal wave. This overestimation of envelope frequency may have been due to the presence of the 210 Hz carrier frequency.

The results of graph comparisons for cases where the starting frequencies were 1 Hz and 300 Hz exhibited a similar trend in frequency perception across different body sites. Frequency perception accuracy was generally low in the forearm but remained consistently high in both the finger pad and the forehead. This suggests that the forehead accurately identifies the frequency due to its ability to detect low-frequency vibrations with minimal noise. Additionally, starting the experimental task at 300 Hz may have influenced the results. Seven of the eleven participants also commented that starting from a higher frequency was more difficult than starting from a lower frequency. The 300 Hz adjustment pattern required pressing the adjustment key longer, making it more difficult than the 1 Hz adjustment. Thus, the observed overestimation of frequency perception at the 300 Hz adjustment pattern is likely attributable to this discrepancy.

Interestingly, the forearm was less accurate in terms of frequency perception than other sites. The superior accuracy of the index finger compared to the forearm may be owing to the higher receptor density of the index finger [33]. Comparing the forearm and forehead, amplitude-modulated vibration signals with a carrier wave of approximately 200 Hz were perceived as noise in the forearm along with the carrier wave itself. On the contrary, the forehead, lacking Pacinian corpuscles sensitive to 200 Hz vibrations, perceived a higher percentage of the envelope wave than the forearm, resulting in better frequency perception accuracy.

The large deviation in the forehead results at 32 Hz, especially when started from 300 Hz, may be owing to the absence of Pacinian corpuscles. Fast-adapting type I (FA I), slow-adapting type I (SA I), and slow-adapting type II (SA II) neurons in the forehead could be involved in frequency perception. Among them, in the forehead, SA I and SA II could be mainly involved in frequency perception because most afferents are slow-adapting neurons [34]. SA I and SA II respond similarly to vibrations below 15 Hz, and their perceived frequency thresholds increase linearly above 20 Hz [35]. This may explain the difficulty in perceiving 32 Hz vibrations on the forehead. Hairy skin, such as the face and forearm, has hair follicle afferents that are sensitive to approximately 30 Hz in the epidermis [36]. Despite the high density of these afferents on the face [37], [38], the experimental results did not suggest that 32 Hz was easily perceived by hair follicle afferents.

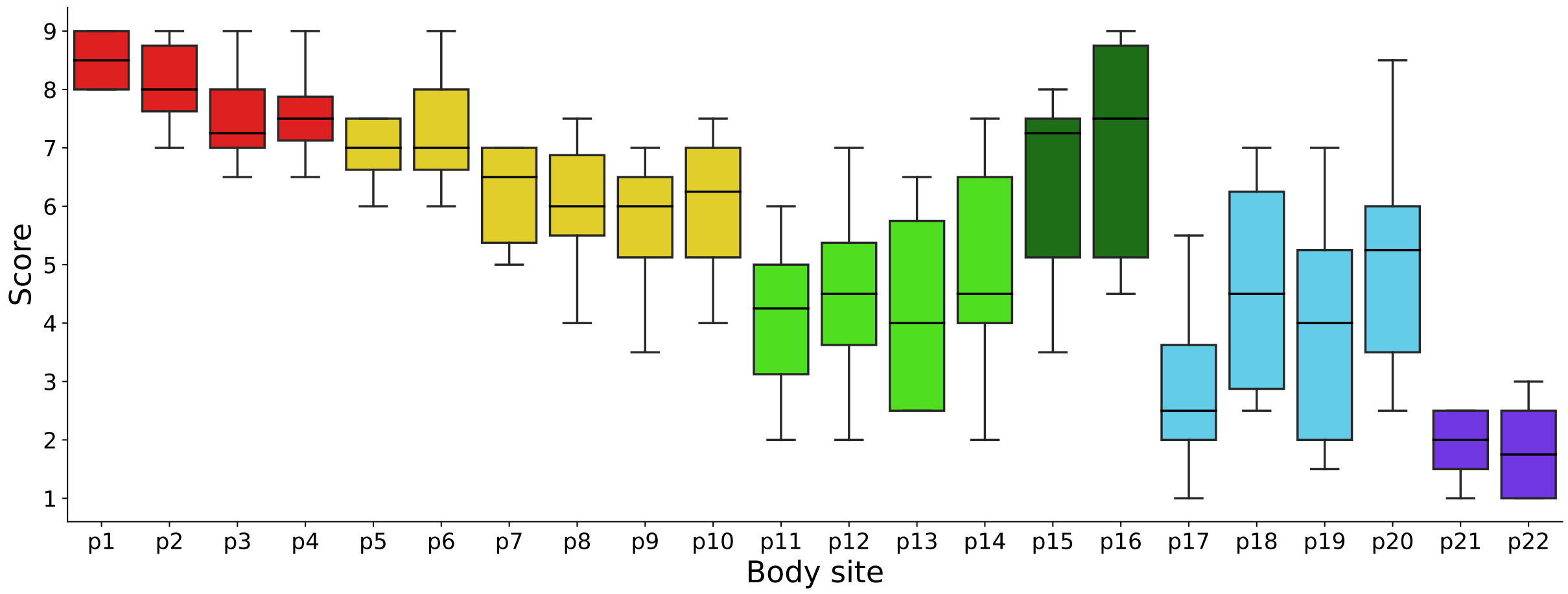

Fig. 11. Results of similarity rating scores for 22 points. The red, yellow, light green-yellow, and green (p1–p16) groups are sites around the face, the light blue group (p17–p20) is around the head, and the purple group is the finger pad and forearm.

It is also possible that bone conduction occurred when the forehead vibrated, and that the frequency was more easily adjusted by the auditory system. However, bone conduction could have been masked by white noise from the headphones. It is also well known that the human forehead has a higher threshold for bone conduction sounds than the condyle and vertex under white noise conditions [39].

### *B. Quality of Tactile Sensation on the Forehead*

In Experiment 2, the forehead consistently scored higher than the finger pad and forearm in terms of smoothness, clarity, and distinctness. This suggests that the forehead is less sensitive to noisy vibrations from amplitude-modulated carrier waves, allowing the envelope wave to be perceived more clearly. Participants' comments highlighted the unique tactile experience of the forehead.

Experiments 1 and 2 showed that the absence of Pacinian corpuscles in the forehead enabled the perception of the envelope of amplitude-modulated vibrations more clearly than at other sites. These findings emphasize the potential of low-frequency stimulation of the forehead through amplitude modulation using a relatively compact LRA.

Based on preliminary studies, we investigated whether pure low-frequency vibrations between 2 Hz and 32 Hz could be presented. However, the range of frequencies in which pure low-frequency vibration is perceived is not known in detail. We believe the limit lies between 16 Hz and 32 Hz. The first reason is that Experiment 1 showed the 32 Hz on the forehead condition gave a large deviation, as shown in Fig. 6 (e). The second reason is that Experiment 2 showed the 32 Hz amplitude-modulated vibration on the forehead did not give different results from the other sites, as shown in Fig. 9.

### *C. Quality of Tactile Sensation on the Face*

In Experiment 3, we identified areas of the face where amplitude-modulated waves produced pure low-frequency sensations, similar to the forehead. The results suggested that the padded area where the HMD was worn (around the eyes), cheeks, and points near the lower lip perceived the amplitude-modulated vibration as pure low-frequency vibration. Many parts of the face produced sensations similar to those of the forehead, likely owing to the absence of Pacinian corpuscles throughout the face [8].

In particular, the region corresponding to the pad of the HMD elicited a sensation similar to that at point pb. This similarity is likely due to the structural resemblance between the skin-bone distance at both locations.

The upper lip part gave low scores, probably due to the mechanoreceptors around the teeth. Levy and Dong [40] suggested that the mechanoreceptors of the teeth and skin may have similar functional mechanisms (with both SA and FA fibers). In addition, the teeth contain periodontal ligament mechanoreceptors (PDLMs) and intradental mechanoreceptors (IMs). Dong et al. [41] reported that PDLMs and IMs are functionally analogous to cutaneous slow-adapting type II mechanoreceptors and Pacinian corpuscles, respectively. These suggest that if the vibration propagated to the teeth, they might have detected high-frequency vibrations and responded that the vibration was noisy. On the contrary, the cheek gave higher scores. Since the cheeks are farther from the teeth, the transmission of vibrations to the teeth might become more difficult. Additionally, the absence of Pacinian corpuscles in the cheeks contributed to the perception of pure low-frequency vibration.

Furthermore, single-fiber recording studies targeting the inferior alveolus have not found evidence of Pacinian corpuscles under the lips and around the jaw [42]. This might explain why these areas also experienced pure low-frequency vibration.

On the other hand, the results in Fig. 11 suggest that the nose, temporal, and parietal areas perceived the presented amplitude-modulated vibration as noisy. Possible explanations for this phenomenon are discussed below.

The temporal and parietal areas of the head may have perceived high-frequency components owing to the propagation of vibrations to the skull. Furthermore, it is possible that Pacinian corpuscles were present in the parietal and temporal regions and perceived high-frequency components as noisy vibrations. We do not have a particular explanation for the low score of the nose part, but conduction through the nasal bone might have influenced the result. As our final goal was to apply low-frequency vibrations that cannot be presented by small vibrators to HMDs, mask-type interfaces, and smart glasses, the results suggest that our method can be applied to the majority of faces.

We believe that this will facilitate the development of two primary applications. The first involves attaching a number of vibrators to the padded portion of the HMD and applying them to applications that require pure low-frequency sensation. For example, the sensation of bubbles jumping into water is better matched by pure sinusoidal low-frequency vibrations than by amplitude-modulated noisy vibrations. This application can be used to create immersive experiences. Second, by incorporating a small vibrator into smart glasses or mask-type interfaces and applying this method, navigation that does not rely on audio-visual perception is possible. This enables a casual tactile presentation. Conventional systems can only present high-frequency vibrations, and if presented to a face without Pacinian corpuscles, it would be difficult to detect the vibrations. However, the proposed method clearly perceived low-frequency vibrations, which may enable directional perception.

### *D. Limitations of the Work*

This study has several limitations. The foremost is our incomplete understanding of the perceptual dynamics for frequencies below 2 Hz. When we tested the 1-Hz amplitude-modulated envelope wave, the resultant forehead vibrations lacked the anticipated smoothness. Although Experiment 1 was conducted by adjusting the frequencies, experiments on frequency discrimination perception should also be conducted to verify the ability to present frequencies.

The proposed method is introduced to deliver pure low-frequency vibrations. Conversely, a method that clearly presents high-frequency vibration to the forehead is also necessary to present material properties. One potential method involves combining auditory and tactile stimuli. Previous research indicates that introducing an unrelated auditory stimulus alongside a tactile stimulus can result in an interaction between the auditory and tactile senses regarding intensity and frequency [43], [44]. This finding suggests that the cross-modal effect between these stimuli may cause low-frequency vibrations to be perceived as high-frequency vibrations.

Furthermore, the mechanism underlying the demodulation of amplitude-modulated vibration signals on the forehead remains elusive. Insufficient contact between the vibrator and the skin could result in a mechanism in which the vibrator intermittently "taps" the skin, where the amplitude modulation signal is demodulated. However, in the present situation, such inadequate contact is unlikely because the actuator was taped to the forehead. Consequently, demodulation likely happened at the receptor or neural activity level or central integration.

## VIII. Conclusion

This study explored the feasibility of presenting low-frequency vibrations across the entire face using amplitude-modulated vibration with a high-frequency carrier wave, aiming for a compact vibrotactile system suitable for integration into wearable devices such as HMDs and smart glasses. We investigated the perceived frequency of low-frequency amplitude-modulated vibrations in the 2-32 Hz range when applied to the finger pad, forearm, and forehead, and asked the participants to evaluate the quality of the vibrations. The results indicated that, within the 2-16 Hz range, the perceived frequency of the forehead was comparable to that of the finger pad and was perceived more accurately than on the forearms. The quality of the vibration on the forehead differed from that of the finger pad and forearm, suggesting that the participants interpreted the amplitude-modulated signal as pure low-frequency vibration. Furthermore, an investigation of the facial areas that evoked similar low-frequency sensations revealed that the regions around the eyes, cheeks, and even the area around the lower lip could produce comparable perceptual effects. As our next step, we intend to apply this approach to wearable interfaces such as HMDs, mask-type interfaces, and smart glasses.

## Acknowledgment

This research was supported by JSPS KAKENHI Grant Number: JP20H05957.

## References

[1] R. Hinchet, V. Vechev, H. Shea, and O. Hilliges, "DextrES: Wearable Haptic Feedback for Grasping in VR via a Thin Form-Factor Electrostatic Brake", in UIST, 2018, pp. 901–912.

[2] I. Choi, E. W. Hawkes, D. L. Christensen, C. J. Ploch, and S. Follmer, "Wolverine: A wearable haptic interface for grasping in virtual reality", in IROS, 2016, pp. 986–993.

[3] Z.-Y. Zhang, H.-X. Chen, S.-H. Wang, and H.-R. Tsai, "ELAXO : Rendering Versatile Resistive Force Feedback for Fingers Grasping and Twisting", in UIST, 2022, pp. 1–14.

[4] T. Kameoka, and H. Kajimoto, "Tactile transfer of finger information through suction tactile sensation in HMDs", In 2021 IEEE WHC, 2021, pp. 949-954.

[5] R. L. Peiris, W. Peng, Z. Chen, L. Chan, and K. Minamizawa, "ThermoVR: Exploring Integrated Thermal Haptic Feedback with Head Mounted Displays", in CHI, 2017, pp. 5452–5456.

[6] V. Shen, C. Shultz, and C. Harrison, "Mouth Haptics in VR using a Headset Ultrasound Phased Array", in CHI, 2022, pp. 1–14.

[7] N. Ranasinghe., et al. "Season Traveller: Multisensory Narration for Enhancing the Virtual Reality Experience", in CHI, 2018, pp. 1–13.

[8] S. M. Barlow, "Mechanical frequency detection thresholds in the human face", Exp Neurol, vol. 96, no. 2, pp. 253–261, 1987, doi: 10.1016/0014-4886(87)90044-6.

[9] Y. Liu et al. "Skin-Integrated Haptic Interfaces Enabled by Scalable Mechanical Actuators for Virtual Reality", IEEE Internet of Things

Journal, vol. 10, no. 1, pp. 653–663, 2023, doi: 10.1109/JIOT.2022.3203417.

[10] M. Manabe, K. Ushiyama, A. Takahashi, and H. Kajimoto, “Energy Efficient Wearable Vibrotactile Transducer Utilizing The Leakage Magnetic Flux of Repelling Magnets”, in IEEE VRW, 2023, pp. 599–600.

[11] H.-Y. Yao, and V. Hayward, “Design and analysis of a recoil-type vibrotactile transducer”, in The Journal of the Acoustical Society of America, vol. 128, no. 2, pp. 619–627, 2010, doi: 10.1121/1.3458852.

[12] J. M. Weisenberger, “Sensitivity to amplitude-modulated vibrotactile signals”, J Acoust Soc Am, vol. 80, no. 6, pp. 1707–1715, 1986, doi: 10.1121/1.394283.

[13] G. Park and S. Choi, “Perceptual space of amplitude-modulated vibrotactile stimuli”, in WHC, 2011, pp. 59-64.

[14] T. Kim, Y. A. Shim, and G. Lee, “Heterogeneous Stroke: Using Unique Vibration Cues to Improve the Wrist-Worn Spatiotemporal Tactile Display”, in CHI, 2021, pp. 1–12.

[15] Y. Akiba, S. Nakayama, K. Ushiyama, I. Mizoguchi, and H. Kajimoto, “Utilizing Absence of Pacinian Corpuscles in the Forehead for Amplitude-Modulated Tactile Presentation”, in EuroHaptics, 2024, pp. 16–28.

[16] S. Choi, and K. J. Kuchenbecker, “Vibrotactile Display: Perception, Technology, and Applications”, in Proceedings of the IEEE, vol. 101, no. 9, pp. 2093–2104, 2013, doi: 10.1109/JPROC.2012.2221071.

[17] N. Cao, H. Nagano, M. Konyo, S. Okamoto, and S. Tadokoro, “A Pilot Study: Introduction of Time-Domain Segment to Intensity-Based Perception Model of High-Frequency Vibration”, in EuroHaptics, 2018, pp. 321–332.

[18] K. Yamaguchi, M. Konyo, and S. Tadokoro, “Sensory Equivalence Conversion of High-Frequency Vibrotactile Signals using Intensity Segment Modulation Method for Enhancing Audiovisual Experience”, in 2021 IEEE WHC, 2021, pp. 674–679.

[19] M. Waga et al. “Representing Fine Texture of Pencil Hardness by High-Frequency Vibrotactile Equivalence Conversion Using Ultra-Thin PZT-MEMS Vibrators”, in IEEE Transactions on Haptics, vol. 17, no. 1, pp. 8–13, 2024, doi: 10.1109/TOH.2023.3349307.

[20] U. A. Alma, R. Rosenkranz, and M. E. Altinsoy, “Perceptual Substitution Based Haptic Texture Rendering for Narrow-Band Reproduction”, in IEEE Transactions on Haptics, vol. 16, no. 2, pp. 204–214, 2023, doi: 10.1109/TOH.2023.3252669.

[21] S. Dosen, A. Ninu, T. Yakimovich, H. Dietl, and D. Farina, “A Novel Method to Generate Amplitude-Frequency Modulated Vibrotactile Stimulation”, in IEEE Transactions on Haptics, vol. 9, no. 1, pp. 3–12, 2016, doi: 10.1109/TOH.2015.2497229.

[22] V. Yem, R. Okazaki, and H. Kajimoto, “Low-Frequency Vibration Actuator Using a DC Motor”, in EuroHaptics, 2016, pp.317-325.

[23] H. Kajimoto, “Forehead Electro-tactile Display for Vision Substitution”, in EuroHaptics, 2006.

[24] H.-Y. Chang, W.-J. Tseng, C.-E. Tsai, H.-Y. Chen, R. L. Peiris, and L. Chan, “FacePush: Introducing Normal Force on Face with Head-Mounted Displays”, in UIST, 2018, pp. 927–935.

[25] J. Gugenheimer, D. Wolf, E. R. Eiriksson, P. Maes, and E. Rukzio, “GyroVR: Simulating Inertia in Virtual Reality using Head Worn Flywheels”, in UIST, 2016, pp. 227–232.

[26] W.-J. Tseng, Y.-C. Lee, R. L. Peiris, and L. Chan, “A Skin-Stroke Display on the Eye-Ring Through Head-Mounted Displays”, in CHI, 2020, pp. 1–13.

[27] V. A. de Jesus Oliveira, L. Brayda, L. Nedel, and A. Maciel, “Designing a Vibrotactile Head-Mounted Display for Spatial Awareness in 3D Spaces”, in IEEE Transactions on Visualization and Computer Graphics, vol. 23, no. 4, pp. 1409–1417, 2017, doi: 10.1109/TVCG.2017.2657238.

[28] D. Valkov and L. Linsen, “Vibro-tactile feedback for real-world awareness in immersive virtual environments”, in 2019 IEEE Conference on Virtual Reality and 3D User Interfaces (VR), 2019, pp. 340-349.

[29] S.-Y. Chu, Y.-T. Cheng, S. C. Lin, Y.-W. Huang, Y. Chen, and M. Y. Chen, “MotionRing: Creating Illusory Tactile Motion around the Head using 360° Vibrotactile Headbands”, in UIST, 2021, pp. 724–731.

[30] I. Hwang and S. Choi, “Perceptual space and adjective rating of sinusoidal vibrations perceived via mobile device”, in 2010 IEEE Haptics Symposium, 2010, pp. 1–8.

[31] G. Park, S. Choi, K. Hwang, S. Kim, J. Sa, and M. Joung, “Tactile effect design and evaluation for virtual buttons on a mobile device touchscreen”, in MobileHCI, 2011, pp. 11–20.

[32] Y. Makino, T. Maeno, and H. Shinoda, “Perceptual characteristic of multi-spectral vibrations beyond the human perceivable frequency range”, in 2011 IEEE WHC, 2011, pp. 439–443.

[33] R. S. Johansson, and A.B. Vallbo, “Tactile sensibility in the human hand: relative and absolute densities of four types of mechanoreceptive units in glabrous skin”, The Journal of Physiology. vol. 286, no. 1, pp. 283-300, 1979, doi: 10.1113/jphysiol.1979.sp012619.

[34] R. S. Johansson, M. Trulsson, K. Å. Olsson, and K.-G. Westberg, “Mechanoreceptor activity from the human face and oral mucosa”, Exp Brain Res, vol. 72, no. 1, pp. 204–208, 1988, doi: 10.1007/BF00248518.

[35] S. Toma, and Y. Nakajima, “Response characteristics of cutaneous mechanoreceptors to vibratory stimuli in human glabrous skin”, Neuroscience Letters, vol. 195, no. 1, pp. 61–63, 1995, doi: 10.1016/0304-3940(95)11776-S.

[36] D. Schmidt, G. Schlee, T. L. Milani, and A. M. C. Germano, “Vertical contact forces affect vibration perception in human hairy skin”, PeerJ. vol. 11, e15952, 2023, doi: 10.7717/peerj.15952.

[37] G. Szabo, “The regional anatomy of the human integument with special reference to the distribution of hair follicles, sweat glands and melanocytes”, Philosophical Transactions of the Royal Society of London. Series B, Biological Sciences, vol. 252, no. 779, pp. 447–485, 1997, doi: 10.1098/rstb.1967.0029.

[38] N. Otberg, H. Richter, H. Schaefer, U. Blume-Peytavi, W. Sterry, and J. Lademann, “Variations of Hair Follicle Size and Distribution in Different Body Sites”, Journal of Investigative Dermatology, vol. 122, no. 1, pp. 14–19, 2004, doi: 10.1046/j.0022-202X.2003.22110.x.

[39] M. McBride, T. Letowski, and P. Tran, “Bone conduction reception: Head sensitivity mapping”, Ergonomics, vol. 51, no. 5, pp. 702–718, 2008, doi: 10.1080/00140130701747509.

[40] J. H. Levy, and W. K. Dong, “Vibration perception thresholds of human vital and nonvital maxillary incisors”, Archives of Oral Biology, vol. 139, 105426, 2022, doi: 10.1016/j.archoralbio.2022.105426.

[41] W. K. Dong, T. Shiwaku, Y. Kawakami, and E. Chudler, “Static and dynamic responses of periodontal ligament mechanoreceptors and intradental mechanoreceptors”, Journal of Neurophysiology, vol. 69, no. 5, pp. 1567–1582, 1993, doi: 10.1152/JN.1993.69.5.1567.

[42] R. S. Johansson and K. A. Olsson, “Microelectrode recordings from human oral mechanoreceptors”, Brain Res, vol. 118, no. 2, pp. 307–311, 1976, doi: 10.1016/0006-8993(76)90716-2.

[43] J. M. Yau, A. I. Weber, and S. J. Bensmaia, “Separate mechanisms for audio-tactile pitch and loudness interactions”, in Frontiers in Psychology, vol.1, 2010, doi: 10.3389/fpsyg.2010.00160.

[44] J. M. Yau, J. B. Olenczak, J. F. Dammann, and S. J. Bensmaia, “Temporal frequency channels are linked across audition and touch”, in Current Biology, vol.19, no.7, pp.561-566, 2009, doi: 10.1016/j.cub.2009.02.013.

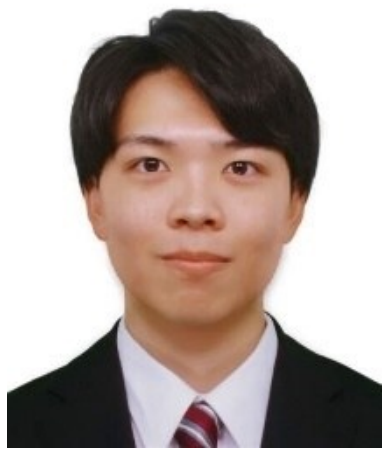

**Yuma Akiba** received a B.E. degree from the University of Electro-Communications, Japan, in 2024, where he is currently pursuing an M.E. degree. His research interests include tactile displays and perception.

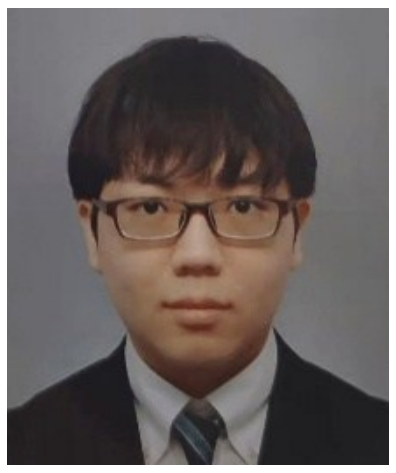

**Shota Nakayama** received B.E. (2022) and M.E. degrees (2024) from the University of Electro-Communications, Japan. His research interests include tactile displays and perception.

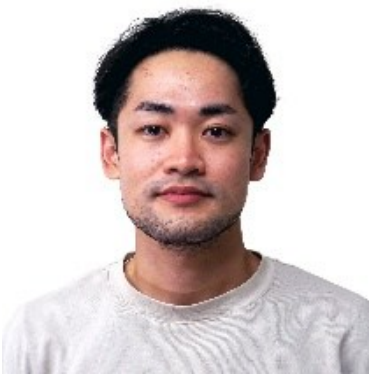

**Keigo Ushiyama** (Member, IEEE) received the Ph.D. degree in Engineering from The University of Electro-Communications in 2025. He is currently a Project Researcher with the Graduate School of Information Science and Technology, The University of Tokyo, Japan. His research interests include tactile displays, proprioceptive displays, and human interfaces.

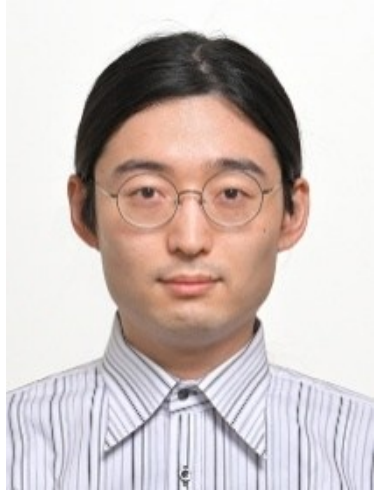

**Izumi Mizoguchi** (Member, IEEE) received a Bachelor of Computer Science degree from the Tokyo University of Technology in 2016, a Master of Engineering degree, and a Ph.D in Engineering in 2018 and 2021 at the University of Electro-Communications. He is currently an assistant professor in the Graduate School of Informatics and Engineering, at the University of Electro-Communications. He researches virtual reality and human augmentation.

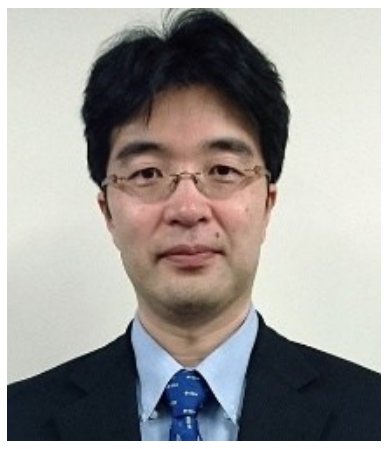

**Hiroyuki Kajimoto** (Member, IEEE) received a Ph.D. degree in information science and technology from the University of Tokyo, in 2006. He is currently a Professor with the Department of Informatics at the University of Electro-Communications, Japan. His current research interests include tactile displays, tactile sensors, human interfaces, and virtual reality.